\documentclass[preprint,aps]{revtex4}

\newcommand {\bc}{\begin{center}}
\newcommand {\ec}{\end{center}}
\newcommand {\bea}{\begin{eqnarray}}
\newcommand {\eea}{\end{eqnarray}}
\newcommand {\be}{\begin{equation}}
\newcommand {\ee}{\end{equation}}

\def\lsim{\mathrel{\rlap{\lower4pt\hbox{\hskip1pt$\sim$}}
    \raise1pt\hbox{$<$}}}               
\def\gsim{\mathrel{\rlap{\lower4pt\hbox{\hskip1pt$\sim$}}
    \raise1pt\hbox{$>$}}}                
\usepackage{graphicx}
\usepackage{color}

\begin{document}


\title{A Generalized Theory of Diffusion based on Kinetic Theory}

\author{T.~Sch\"afer}

\affiliation{Department of Physics, North Carolina State University,
Raleigh, NC 27695}

\begin{abstract}
We propose to use spin hydrodynamics, a two-fluid model of spin 
propagation, as a generalization of the diffusion equation. We show 
that in the dense limit spin hydrodynamics reduces to Fick's law
and the diffusion equation. In the opposite limit spin hydrodynamics 
is equivalent to a collisionless Boltzmann treatment of spin 
propagation. Spin hydrodynamics avoids unphysical effects that arise 
when the diffusion equation is used to describe to a strongly 
interacting gas with a dilute corona. We apply spin hydrodynamics 
to the problem of spin diffusion in a trapped atomic gas. We find
that the observed spin relaxation rate in the high temperature limit 
[Sommer et al., Nature 472, 201 (2011)] is consistent with the 
diffusion constant predicted by kinetic theory. 

\end{abstract}

\maketitle

\section{Introduction}
\label{sec_intro}

 Diffusion plays an important role in many areas of physics, and the 
problem of finding numerical and analytical solutions to the diffusion 
equation is well understood \cite{Crank:1980}. However, many interesting 
applications of the diffusion equation involve problems in which the mean 
free path varies significantly, so that the diffusion approximation breaks 
down in the dilute, weakly collisional, regime. In this case a naive treatment
of the diffusion equation will lead to unphysical results. In a dilute 
gas the diffusion coefficient scales inversely with the density, and 
the diffusion current can become unphysically large. This problem can 
be dealt with in a phenomenological way by using flux limiters or boundary 
conditions. However, given that the dilute regime is physically well 
understood, it should be possible to derive quantitatively accurate 
schemes that interpolate between diffusion and ballistic motion. 

 In this work we propose a generalization of the diffusion equations 
that correctly extrapolates to the ballistic limit. The method is 
based on moments of the Boltzmann equation, and bears some resemblance 
to moment methods employed for radiation hydrodynamics in astrophysics 
\cite{Pons:2000,OConnor:2015}. The method was inspired by recent work 
on anisotropic fluid dynamics, which has been used to implement the 
correct ballistic limit of the Navier-Stokes equation in relativistic 
and non-relativistic fluid dynamics 
\cite{Florkowski:2010,Martinez:2010sc,Bluhm:2015raa,Bluhm:2015bzi} 
(see \cite{Brewer:2015hua} for a different approach to this problem,
based on the lattice Boltzmann method).

 The work was motivated by attempts to extract the spin diffusion
constant of ultracold atomic gases from experiments with optically
trapped atoms \cite{Sommer:2011,Koschorreck:2013,Valtolina:2016}, see 
also \cite{Du:2008,Du:2009,Badon:2014,Trotzky:2015}. A particularly 
interesting system is the two-component unitary Fermi gas. In this 
case the two-body scattering length is infinite, and the diffusion 
constant is expected to enter the quantum regime $D\sim \hbar/m$, 
where $m$ is the mass of the particles \cite{Bruun:2010}. The 
determination of the spin diffusion constant from experiment is in 
principle straightforward. The experiment involves preparing a 50-50 
mixture of spin up and down particles. The two spin components are 
spatially separated and then released. The early time dynamics is 
typically complicated, but at late times exponential relaxation to 
a locally balanced mixture is observed. The diffusion constant depends 
on the local density $n$ and temperature $T$, but this dependence can 
be unfolded by performing experiments at different temperatures, and 
for different numbers of particles. In the unitary Fermi gas the 
situation is further simplified by scale invariance, which implies 
that $D=\frac{\hbar}{m}f(mTn^{-2/3})$ where $f(x)$ is a function of a 
single variable. 

 The tool for extracting the diffusion constant is the diffusion equation. 
We have to construct solutions of the diffusion equation in a given trap 
geometry and adjust the diffusion constant in order to achieve agreement 
with the observed spin relaxation times. The difficulty, as pointed out 
in the present context by Bruun and Pethick \cite{Bruun:2011b}, is that 
the diffusion approximation breaks down in the dilute part of the cloud. 
If this issue is ignored, observed spin relaxation times disagree with 
theoretical expectations by more than an order of magnitude. Bruun and
Pethick proposed to address this issue by imposing a transverse cutoff
on the diffusion equation in an elongated trap. The cutoff radius is
determined by a simple mean free path estimate, or fitted to experiment.
A similar procedure for estimating shear viscosity was used in 
\cite{Joseph:2014}. 

 In the present work we propose to improve on this procedure by deriving a 
generalization of the diffusion equation which we call ``spin hydrodynamics''.
Spin hydrodynamics describes the transition from diffusive to ballistic
behavior dynamically, based on a relaxation time equation. The paper is 
structured as follows. In Sect.~\ref{sec_diff} we review the derivation 
of Fick's law from kinetic theory, and in Sect.~\ref{sec_sol} we discuss 
the behavior of variational and numeric solutions of the diffusion equation 
in a harmonically trapped gas. The equations of spin fluid dynamics are 
derived in Sect.~\ref{sec_shydro}, and the diffusive and ballistic limits 
are studied in Sect.~\ref{sec_lim}. A numerical method for implementing 
spin hydrodynamics is described in Sect.~\ref{sec_sim}. Numerical tests 
are presented in Sect.~\ref{sec_num}, and numerical results in a trap 
geometry are given in Sect.~\ref{sec_num_trap}. We provide an outlook
in Sect.~\ref{sec_out}.

\section{Kinetic theory and the diffusion equation}
\label{sec_diff}

 In this section we review the derivation of the spin diffusion
equation from kinetic theory in a two-component Fermi gas. Consider
the Boltzmann equation
\be 
\label{BE}
\left( \partial_0 + \vec{v}\cdot\vec{\nabla}_x + \vec{F}\cdot\vec\nabla_p
  \right) f_{p\sigma}(x,t) = C[f_{p\sigma}] \, , 
\ee
where $f_{p\sigma}(x,t)$ is the phase space density of particles with spin 
$\sigma=\uparrow\downarrow$, $\vec{v}$ is the velocity of the particles, 
$\vec{F}$ is a force, and $C[f_{p\sigma}]$ is the collision term. For
quasi-particles with energy $E_p$ we have 
\be 
 \vec{v}= \vec\nabla_p E_p\, ,\hspace{1cm}
 \vec{F}=-\vec\nabla_x E_p\, . 
\ee
We will focus on the case $E_p=\epsilon_p+V(x)$, where $\epsilon_p$ is 
solely a function of momentum, and $V(x)$ is an external spin-independent
potential. We are interested in the spin current $\vec\jmath_M=\vec
\jmath_\uparrow-\vec\jmath_\downarrow$ generated in response to a magnetization 
gradient $\vec\nabla M$, where $M=n_\uparrow-n_\downarrow$. Here, the 
spin densities and currents are given by
\be 
 n_\sigma(x,t) = \int d\Gamma\, f_{p\sigma}(x,t)\, , \hspace{1cm}
 \vec\jmath_\sigma(x,t) = \int d\Gamma\, \vec{v}\,f_{p\sigma}(x,t)\, ,
\ee
where $d\Gamma=d^3p/(2\pi)^3$. If the collision term conserves spin
then the Boltzmann equation implies 
\be 
\label{spin_cons}
\partial_0 M + \vec\nabla \cdot \vec\jmath_M = 0\, . 
\ee
We will focus on near-equilibrium distributions of the form
\bea
\label{Chap_Ens}
 f_{p\sigma}(x,t)&=&f^0_{p\sigma}(x,t)\left( 1 +\frac{\chi_{p\sigma}(x,t)}{T}
   \right)\, , \\
\label{f_0}
 f^0_{p\sigma}(x,t)&=&\exp\left(-\frac{1}{T(x,t)}
    \left[\epsilon_p+V(x)-\mu_\sigma(x,t)\right]\right)\, .
\eea
For simplicity we make the relaxation time (Bhatnagar-Gross-Krook, BGK)
approximation to the collision term
\be 
\label{BGK}
C[f_{p\sigma}] = -\frac{f^0_{p\sigma}\chi_{p\sigma}}{T\tau}\, , 
\ee
where $\tau$ is a collision time. It is straightforward to solve the 
Boltzmann equation at leading order in $\tau$ and in gradients of 
the thermodynamic variables. We find 
\be 
\label{chi_BGK}
\chi_{\sigma p} = -\tau \vec{v}\cdot\vec\nabla \mu_\sigma
\ee
and 
\be 
\label{Fick_1}
\vec\jmath_M = -D_\mu \vec\nabla \delta\mu\, , \hspace{1cm}
 D_\mu = \frac{\tau}{3T}\int d\Gamma\, v^2 f^0_p\, ,
\ee
where $\delta\mu=\mu_\uparrow-\mu_\downarrow$. For $\epsilon_p=p^2/(2m)$ 
we get $D_\mu=(\tau n)/(2m)$. Finally, we obtain the standard form
of Fick's law by changing variables from $\delta\mu$ to $M$, 
\be 
\label{Fick_2}
\vec\jmath_M = -D \left[ \vec{\nabla} M
    -  k_n \vec{\nabla} n \right]\, , \hspace{1cm} 
 D=\chi^{-1}_M D_\mu\, ,
\ee
where $\chi_M = (\partial M)/(\partial \delta \mu)$ and $k_n = \chi_n/
\chi_M$ with $\chi_n = (\partial n)/(\partial \delta\mu)$. For a 
non-interacting gas $\chi_M=n/(2T)$, $k_n=M/n$ and $D=(\tau T)/m$. 
Note that $D$ has units $\hbar/m$, and the quantum limit corresponds 
to $\tau\sim\hbar/T$. In the following we will set $\hbar=k_B=1$. For 
a given collision term we can express the collision time $\tau$ in
terms of the scattering parameters. In the dilute Fermi gas at unitarity 
we have $\sigma=4\pi/k^2$ where $k$ is the relative momentum of the 
spin up and down particles. Solving the Boltzmann equation at leading
order in gradients gives \cite{Bruun:2010,Sommer:2011}
\be 
\label{D_CE}
 D = \frac{9\pi^{3/2}}{32\sqrt{2}m} \left( \frac{T}{T_F}\right)^{3/2} \, ,
\ee
where $T_F=k_F^2/(2m)$ is the Fermi temperature, and $k_F=(3\pi^2 n)^{1/3}$ 
is the Fermi momentum. The result in equ.~(\ref{D_CE}) was obtained at leading 
order in an expansion of $\chi_{\sigma p}$ in Laguerre polynomials. The next 
order correction has not been computed, but the corresponding approximation 
is known to be accurate to better than $2\%$ for other transport coefficients, 
such as the shear viscosity. The most important feature of equ.~(\ref{D_CE}) 
is that $D\sim 1/n$, which is a general result that follows from kinetic 
theory in the dilute limit. More detailed studies of spin diffusion were 
performed by Enss and collaborators 
\cite{Enss:2012,Enss:2012b,Enss:2013,Enss:2015}.

\section{Diffusion in the high and low temperature limits}
\label{sec_sol}

 Solutions to the diffusion equation is a trapped atomic system were 
studied by Bruun and Pethick \cite{Bruun:2011b}. Here we will briefly 
review their study, and generalize the result to low temperature gases. 
We consider the diffusion equation, equ.~(\ref{spin_cons}) and (\ref{Fick_2}). 
We will assume $k_n=M/n$, so that the diffusion equation takes a simple
form when written in terms of the polarization $P=M/n$. We find
\be
\partial_0 P - \frac{1}{n}\vec\nabla\left[ nD\,\vec\nabla P\right] = 0\, . 
\ee
We are interested in solutions of the form $P(x,t)=e^{-\Gamma_i t}P_i(x)$. 
In the asymptotic limit the solution is dominated by the lowest mode 
$\Gamma\equiv\Gamma_0$. This equation further simplifies in the high
temperature  limit where $nD={\it const}$. In that case the diffusion 
equation is 
\be 
\partial_0P - \frac{n(0)D(0)}{n}\nabla^2 P = 0\, ,
\ee 
where $n(0)$ and $D(0)$ are the density and diffusion constant at the 
trap center. Bruun and Pethick observed that this equation can be solved 
using variational methods, in analogy to the Schr\"odinger equation. The 
variational bound on $\Gamma$ is 
\be 
\label{Gamma_var}
 \Gamma \leq n(0)D(0)\, \frac{\displaystyle\int d^3x\, [\vec\nabla P_v(x)]^2}
                            {\displaystyle\int d^3x\, n(x) P_v(x)^2}\, , 
\ee 
where $P_v(x)$ is a variational function. Consider a dilute Fermi gas 
in a harmonic trapping potential $V(x)=\frac{1}{2}m\omega_i^2x_i^2$. In 
that case $n(x)=n(0)\exp(-V(x)/T)$. We will focus on axially symmetric
potentials $\omega_x=\omega_y\equiv \omega_\perp$ and $\omega_z=\lambda
\omega_\perp$. On dimensional grounds we have
\be 
\label{Gamma_dim}
 \Gamma = \frac{D(0)}{l_z^2}\Gamma_{\it red}(\lambda)\, , 
\ee
where $l_z^2=2T/(m\omega_z^2)$ is the square of the oscillator length 
in the $z$-direction, and $\Gamma_{\it red}$ is a dimensionless damping
constant. A variational ansatz with the correct symmetry and asymptotic
behavior is 
\be 
 P_v(x) = \frac{z}{1+\tilde{R}^3}\, , \hspace{1cm}
 \tilde{R} = \left(\frac{x^2+y^2}{d_\rho^2}+\frac{z^2}{d_z^2}\right)^{1/2}\, , 
\ee
where $d_\rho$ and $d_z$ are variational parameters. Using this ansatz we 
find $\Gamma_{\it red}(\lambda\!=\!0)= 12.1$, $\Gamma_{\it red}(\lambda\!=\!0.4)
= 29.2$ and $\Gamma_{\it red}(\lambda\to 0) = \lambda^{-2}/\log(0.13\lambda^{-2}
)$. The limit $\lambda\to 0$ can be derived rigorously using a WKB 
approximation. 

 The experimental work reports the spin drag coefficient $\Gamma_{sd}=
\omega_z^2/\Gamma$ in units of the Fermi Energy $E_F(0)$. Note that 
$E_F(0)$ refers to the local Fermi energy at the trap center. The result 
is based on the observed decay rate of the spin dipole moment. In the high 
temperature limit Sommer et al.~find $\Gamma_{sd}=0.16\, E_F(0)(T_F/T)^{1/2}$ 
\cite{Sommer:2011}. The experimental paper does not provide the value 
of $\lambda$, but states that in the regime that was investigated the 
spin drag $\Gamma_{sd}/E_F(0)$ is independent of $\lambda$. Using 
equ.~(\ref{D_CE}) and equ.~(\ref{Gamma_dim}) we obtain the theoretical
prediction
\be 
\label{G_sd}
 \Gamma_{sd}= \frac{1.81\,E_F(0)}{\Gamma_{\it red}(\lambda)}
  \left(\frac{T_F}{T}\right)^{1/2}\, . 
\ee
For a strongly deformed cloud $\Gamma_{\it red}\gsim \Gamma_{\it red}(0.1)\simeq 
200$, which differs from the experimental result $\Gamma_{\it red}\simeq 
11.3$ by more than an order of magnitude. Bruun and Pethick argued that the 
discrepancy is related to the treatment of the dilute part of the cloud, and 
suggested imposing a transverse cutoff $r_0$ in equ.~(\ref{Gamma_var}). The 
result is very sensitive to the precise value of $r_0$, but the experimental 
result can be understood for a reasonable value $r_0=2.1l_\perp$, where 
$l_\perp$ is the transverse oscillator length. 

 For comparison we have studied diffusion in a low temperature gas. Here, 
we assume that the low temperature limit corresponds to $D=D(0)$, which means 
that the diffusion constant is only a function of temperature and not of 
density. This is a slight idealization, because in a degenerate Fermi gas 
the diffusion constant is expected to exhibit the Landau Fermi liquid 
behavior $mD\sim (T_F/T)^2$ \cite{Bruun:2010}. Combined with equ.~(\ref{D_CE}) 
this result implies that $mD$ has a minimum as a function of $T/T_F$, and
that near the minimum there is a regime in which the diffusion constant is 
approximately density independent. In this limit the diffusion equation is
\be 
\partial_0P - \frac{D(0)}{n}\vec\nabla\left[ n\vec\nabla P\right] =0 \, .
\ee
The variational principle gives
\be 
\label{Gamma_var_2}
 \Gamma \leq D(0)\, \frac{\displaystyle\int d^3x\, n(x)[\vec\nabla P_v(x)]^2}
                            {\displaystyle\int d^3x\, n(x) P_v(x)^2}\, .
\ee 
This equation is minimized by $\Gamma_{\it red}=2$ and $P_v(x)\sim z$, 
independent of $\lambda$. The result that $\Gamma_{\it red}$ is 
approximately $\lambda$-independent is consistent with experiment, 
but the value of $\Gamma_{\it red}$ is not. Whereas the value $\Gamma_{\it red}$ 
in the dilute limit is too  large, the value in the dense limit is too 
small. This suggest that the correct spin current profile must be 
intermediate between the structure in the high and low temperature limits.  

\begin{figure}[t]
\bc\includegraphics[width=7.5cm]{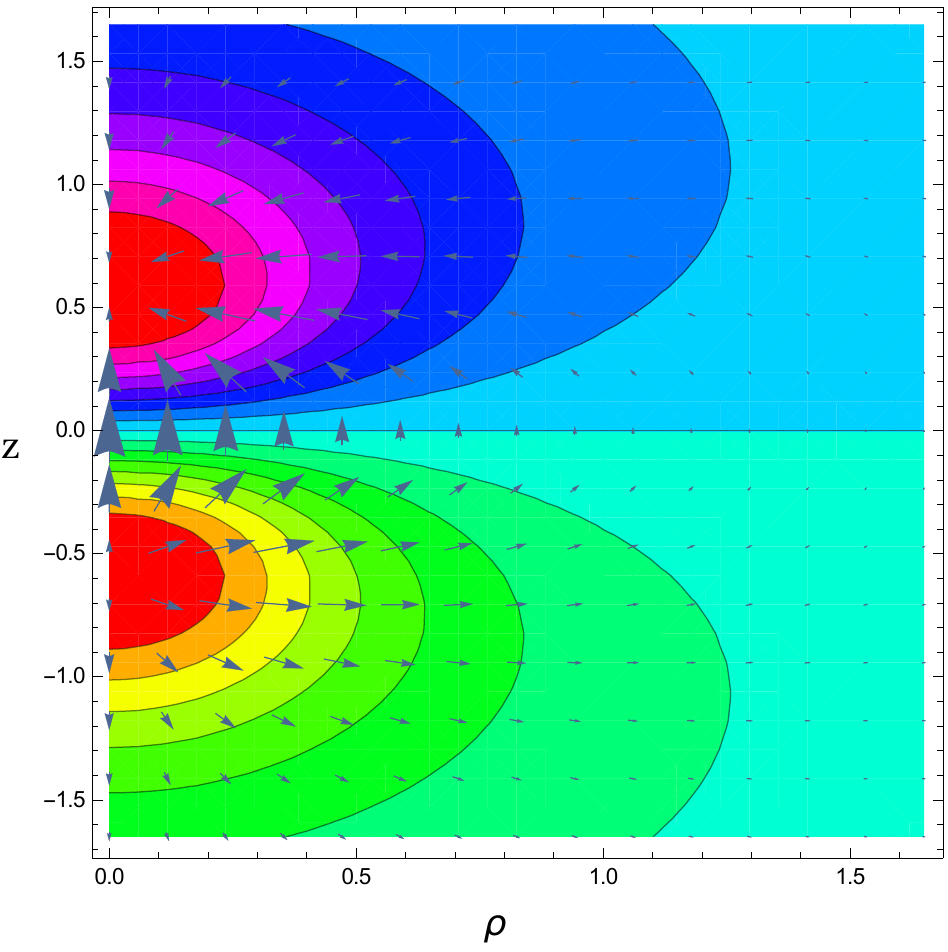}
\hspace*{0.2cm}
\includegraphics[width=7.3cm]{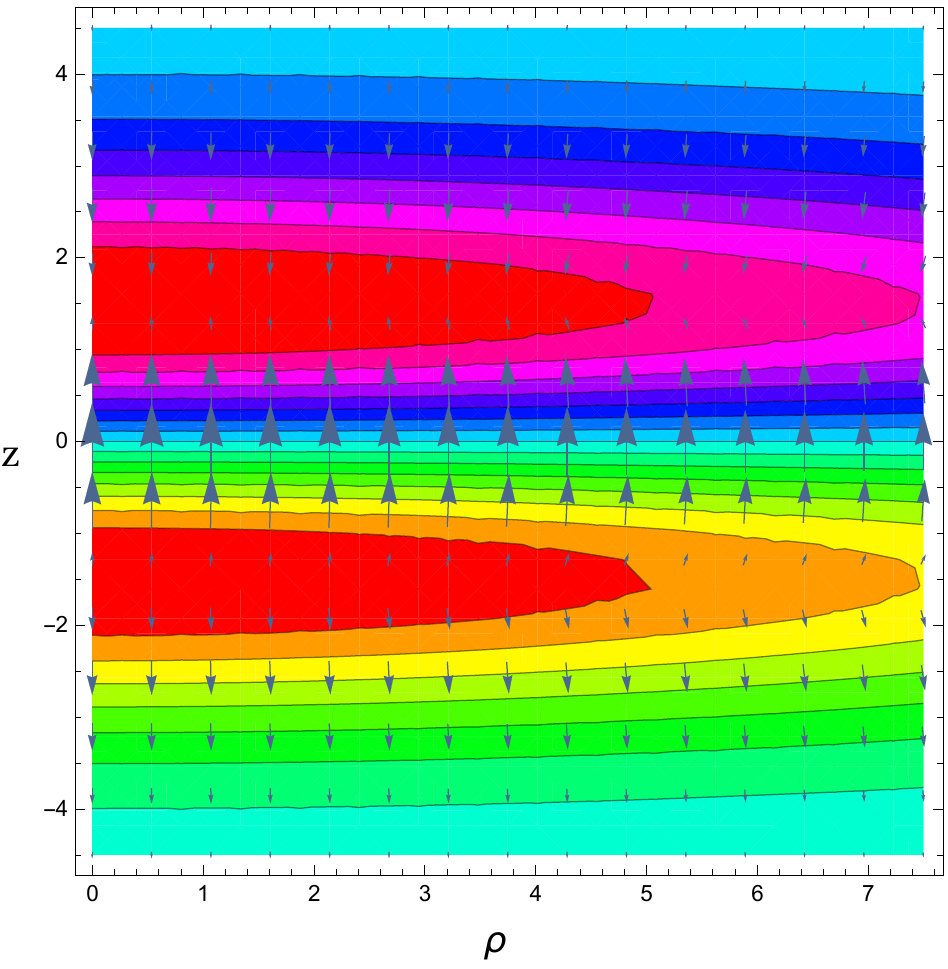}\ec
\caption{\label{fig_dif}
Solutions of the spin diffusion equation for a gas confined in a 
harmonic potential with deformation $\lambda=0.4$. The contours show 
the polarization $P$ as a function of the dimensionless variables 
$\bar\rho$ and $\bar{z}$, and the vector field shows the spin current 
$\vec\jmath$. The contour plots have 15 equally spaced contour lines
between the maximum and minimum polarization at the center of the 
trap. The left panel shows a solution in the high temperature 
limit $D=D(0)n(0)/n$, and the right panel corresponds to the low 
temperature limit $D=D(0)$.  }   
\end{figure}

 In order to verify the variational estimates we have numerically solved
the diffusion equation in the high and low temperature limits. In the 
high temperature limit we assume that $D=D(0)n(0)/n$. The diffusion equation 
in cylindrical coordinates is 
\be
\label{diff_cyl_1} 
\partial_{\bar{t}} P - e^{-\bar{V}}\left[ 
  \frac{1}{\bar\rho} \partial_{\bar\rho} 
       \left( \bar{\rho}\partial_{\bar\rho} P\right) 
    + \partial_{\bar{z}}^2 P \right] = 0\, , 
\ee
where $\bar\rho=(x^2+y^2)^{1/2}/l_z$ and $\bar{z}=z/l_z$ are 
dimensionless variables and $\bar{V}=\lambda^{-2}\bar{\rho}^2+\bar{z}^2$.
The dimensionless time variable is $\bar{t}=m\omega_z^2D(0)t/(2T)$, so 
that $\Gamma$ is automatically given in units of $D(0)/l_z^2$. A solution
of the diffusion equation for $\lambda=0.4$ is shown in Fig.~\ref{fig_dif}.
The decay constant of the spin current is $\Gamma_{\it red}\simeq 29$ which 
agrees with the variational estimate $\Gamma_{\it red}=29.2$. It is important 
to note that the spin current is not quasi one-dimensional, even in a deformed 
trap. 

 Using cylindrical coordinates the diffusion equation in the dense limit
is given by 
\be 
\label{diff_cyl_2}
\partial_{\bar{t}} P -  \left[
     \partial_{\bar\rho}^2  
   + \frac{1}{\bar{\rho}}\partial_{\bar\rho} 
   + \partial_{\bar{z}}^2 
   - 2\left(  \bar{z}\partial_{\bar{z}} 
             + \frac{\bar\rho}{\lambda^2}\partial_{\bar\rho} \right)
  \right] P  = 0\, .
\ee
A solution of the diffusion equation is shown in the right panel of 
Fig.~\ref{fig_dif}. We observe that the distribution of spin current 
is very different from the dilute limit. In particular, we find that 
diffusion is approximately one-dimensional. The decay constant is 
$\Gamma_{\it red}\simeq 2$, in very good agreement with the variational 
estimate. This result implies that the decay of the magnetization
is much slower (by almost a factor 15) as compared to the dilute 
limit. This result is easy to understand: In the dilute regime 
spin polarization decays by generating a large spin current in the 
dilute corona. In the dense limit the polarization has to decay 
by producing much smaller currents in the dense part of the cloud.

\section{Spin hydrodynamics and kinetic theory}
\label{sec_shydro}

 In order to improve the accuracy of the diffusion equation in the 
dilute limit we revisit the derivation of the diffusion equation in
kinetic theory. Consider the Boltzmann transport equation, equ.~(\ref{BE}), 
with a two-body collision term
\be 
\label{coll_2bdy}
C[f_{p_1\sigma_1}] = \sum_{\sigma_2\sigma_3\sigma_4}\int d\Gamma_{234}\, 
 \left( f_{p_1\sigma_1}f_{p_2\sigma_2} - f_{p_3\sigma_3}f_{p_4\sigma_4}\right)
     w(p_1\sigma_1,p_2\sigma_2; p_3\sigma_3,p_4\sigma_4)\, , 
\ee 
where $w$ is the transition amplitude. We assume that $w$ is of the
form 
\be 
  w(p_1\sigma_1,p_2\sigma_2; p_3\sigma_3,p_4\sigma_4)
 = (2\pi)^4 \delta\big( \sum_i E_i\big)
            \delta\big( \sum_i p_i\big)
            \delta_{\sigma_1+\sigma_2,\sigma_3+\sigma_4}
            |{\cal A}_{\sigma_1\sigma_2}(P,q)|^2
\ee
where $2P=p_1+p_2$ and $2q=p_1-p_2$. In this case moments of the 
collision operator with respect to particle number, momentum, and
energy vanish
\be 
\label{C_mom}
 \sum_\sigma\int d\Gamma\, R_i(p) C[f_{p\sigma}] = 0 \, , 
\ee
where $R_i=\{1,\vec{p},\epsilon_p\}$. Similarly, conservation of 
spin implies 
\be 
\label{C_mom_2}
 \sum_\sigma\int d\Gamma\, \bar\sigma C[f_{p\sigma}] = 0 \, , 
\ee
where $\bar\sigma=\pm$ for $\sigma=\uparrow,\downarrow$. This relation does 
not generalize to other moments such as $\bar\sigma\vec{p}$ and $\bar\sigma
\epsilon_p$. The Boltzmann equation and equ.~(\ref{C_mom}) imply conservation 
laws for particle number, momentum, and energy
\bea
\label{hydro_0}
\partial_0 n + \vec\nabla\cdot\vec\jmath_n &=&0\, , \\ 
\label{hydro_1}
\partial_0\pi^i + \nabla_j\Pi^{ij} &=&0\, , \\
\label{hydro_2}
\partial_0{\cal E} + \vec\nabla\cdot\vec\jmath_\epsilon&=&0\, .
\eea
Here, $n=n_\uparrow+n_\downarrow$, $\vec\jmath_n=\vec\jmath_\uparrow + 
\vec\jmath_\downarrow$ and $\vec{\pi}=m\vec\jmath_n$. We also have 
\bea 
\Pi_{ij}    &=& \sum_\sigma \int d\Gamma\, f_{\sigma p}\, p_iv_j\, , \\
{\cal E}\, &=& \sum_\sigma \int d\Gamma\, f_{\sigma p}\, \epsilon_p \, , \\
\vec\jmath_\epsilon\, &=& \sum_\sigma \int d\Gamma\, 
  f_{\sigma p}\, \vec{v}\,\epsilon_p \, .
\eea
Equ.~(\ref{C_mom_2}) implies the spin conservation equation (\ref{spin_cons}).
In order to derive the diffusion equation we need a constitutive equation
for the spin current $\vec\jmath_M$. As shown in Sect.~\ref{sec_diff} Fick's
law $\vec\jmath_M=-D\vec{\nabla}M$ can be derived by assuming that $f_{p\sigma}$
is close to the equilibrium distribution, see equ.~(\ref{Chap_Ens}). In 
this section we will follow a different strategy. We derive an equation
of motion for $\vec\jmath_\sigma$ from the $\vec{p}$ moment of 
the Boltzmann equation for each $\sigma$. We find
\be 
 \partial_0 (m\jmath_\sigma^i)
 + \nabla_j \Pi^{ij}_\sigma - F^i n_\sigma
 = \int d\Gamma\, p^i \, C[f_{p\sigma}]\, . 
\ee
In order for the equations of motion to close we need a constitutive
equation for the spin stress $\Pi^{ij}_\sigma$, and an explicit expression
for the collision term. We will make a generalized ansatz for the 
distribution function
\be 
\label{f_shydro}
 f_{p\sigma}(x,t) =\exp\left(\frac{1}{T(x,t)}\left[
  \mu_\sigma(x,t) -\frac{1}{2m}\left(p^i-mu^i_\sigma(x,t)\right)^2
      \right]\right)\, ,
\ee
where $\vec{u}_\sigma$ is a spin velocity. Note that this distribution
functions includes the Chapman-Enskog ansatz in equ.~(\ref{Chap_Ens}) and 
(\ref{chi_BGK}) as a special case. If $\vec{w}=\frac{1}{2}(\vec{u}_\uparrow-
\vec{u}_\downarrow)$ is small we can expand equ.~(\ref{f_shydro}) and
obtain
\be 
 f_{p\sigma}(x,t) \simeq f^0_{p\sigma}(x,t) \left( 
  1 \pm \frac{m}{T} \vec{v}\cdot\vec{w} \right)\, , 
\ee
where the $\pm$ sign corresponds to $\sigma=\uparrow\downarrow$.  We 
observe that equ.~(\ref{chi_BGK}) is recovered for $m\vec{w}=-\frac{\tau}{2}
\vec{\nabla}\delta\mu$. However, if $\vec{w}$ is large then $f_{p\sigma}$ 
is not close to equilibrium. We will show below that equ.~(\ref{f_shydro}) 
solves the Boltzmann equation in the ballistic limit, and in this way 
provides a smooth connection between the diffusive and ballistic limits. 

 We can now derive equations of motion by taking moments of the Boltzmann 
equation with respect to particle number and momentum for fixed spin. Moments 
with respect to particle number give the continuity equations
\be 
\label{shydro_0}
\partial_0 n_\sigma + \vec\nabla\cdot \left( n_\sigma\vec{u}_\sigma\right)
  = 0 \, . 
\ee
Moments with $\vec{p}$ give equations of motion for $n_\sigma\vec{u}_\sigma$. 
We get 
\be 
\label{shydro_1}
\partial_0 \left( mn_\sigma u^i_\sigma \right)
  + \nabla_j \Pi_\sigma^{ij} + n_\sigma F^i = S_\sigma   \, , 
\ee
where $F^i$ is an external force and we have defined the spin stresses
\be 
\label{shydro_2}
 \Pi_\sigma^{ij} = mn_\sigma u_\sigma^i u_\sigma^j 
    + n_\sigma T \delta^{ij} \, . 
\ee 
The source term $S_\sigma$ depends on the collision term. In the BGK 
approximation
\be 
 C[f_{p\sigma}] = -\frac{f_{p\sigma}-f^0_{p\sigma}}{\tau}\, , 
\ee
where $f^0_{p\sigma}$ is given in equ.~(\ref{f_0}) we obtain
$S_\sigma =  \mp(mn_\sigma w^i)/\tau$. This result exhibits some unphysical 
features, related to shortcomings of the BGK approximation. In particular, 
$S_\sigma$ does not conserve the total momentum of spin up and down particles, 
even though the microscopic collision term in equ.~(\ref{coll_2bdy}) conserves 
momentum. We address this problem by replacing $n_\sigma\to n_g$, where 
$n_g=n_\uparrow n_\downarrow/(n_\uparrow+n_\downarrow)$ is the geometric mean
of the up and down densities. This gives 
\be 
\label{S_sigma}
S_\sigma =  \mp\frac{mn_gw^i}{\tau}\, . 
\ee
Like the BGK collision term, this is a model for collisional relaxation
in a two component gas. It does, however, have two advantages compared
to the BGK model: i) It conserves total momentum; ii) The collision rate 
goes to zero if either one of the two densities goes to zero, as predicted 
by the full collision term. We note that the collision term is characterized 
by a single parameter $\tau$, which may depend on $n$ and $T$. In the 
following section we will show that in order to reproduce the diffusion
equation with diffusion constant $D(n,T)$ the relaxation time should be 
chosen as
\be 
\label{tau_dif}
 \tau(T,n) = \frac{mD(n,T)}{T}\, . 
\ee
In a weakly polarized gas ($n_\uparrow\simeq n_\downarrow$) this is the 
same relation we obtained from the BGK model in Sect.~\ref{sec_diff}.

Equ.~(\ref{shydro_0}-\ref{shydro_2}) are the defining equations of 
spin hydrodynamics. We note that the equations indeed close. There 
are eight variables $n_\uparrow,n_\downarrow,\vec{u}_\uparrow$ and 
$\vec{u}_\downarrow$ and eight equations of motion. This is the case
as long as we consider the temperature of the cloud to be fixed. 
If the evolution of $T$ is needed then we can add an equation for 
the total energy density ${\cal E}$, see equ.~(\ref{hydro_2}).
We also note that if $\vec{u}\equiv \vec{u}_\uparrow=\vec{u}_\downarrow$
summing equ.~(\ref{shydro_0}-\ref{shydro_2}) gives the usual 
Euler equation. If viscous effects are important, then we can
either extend equ.~(\ref{f_shydro}) to include an anisotropic 
temperature as in \cite{Bluhm:2015raa}, or include a spin-independent
term in $\Pi^{ij}_\sigma$ which is proportional to the viscous stresses. 
 
\section{Diffusive and ballistic limits}
\label{sec_lim}

 In this section we will check that spin hydrodynamics does indeed
correctly reproduce the diffusive and ballistic limits. First 
consider the diffusive case. The difference of the continuity 
equations gives
\be 
 \partial_0 M + \vec\nabla\cdot \left( M\vec{u}+n\vec{w} \right) =0\, . 
\ee
The first term in the spin current is the advection term $\vec\jmath_M
\sim M\vec{u}$. The second term, $\vec\jmath_M\sim n\vec{w}$ can be 
computed using the difference of the spin stress equations. In the 
diffusive limit these equations can be solved order by order in the 
small parameter $\tau T$. At leading order, and ignoring external 
forces, we find $\vec{w}=-\frac{\tau T}{mn} \vec\nabla M+\vec{w}_a$. 
Here, $\vec{w}_a$ is an $O(\tau)$ correction to the advection term
$M\vec{u}$. Neglecting this term, we get
\be 
 \partial_0 M - \vec\nabla\cdot \left( D\vec\nabla M-\vec{u}M\right)=0 \, ,
\ee
with $D=\tau T/m$, in agreement with the result in Sect.~\ref{sec_diff}.
We can also study the effect of an external force. In hydrostatic 
equilibrium we neglect the time derivatives and velocity terms. We
get 
\be 
 \frac{T\vec\nabla n_\sigma}{n_\sigma} = - \vec\nabla V_{\it ext}\, , 
\ee
which implies $n_\sigma(x) \sim \exp(-V_{\it ext}(x)/T)$. We can use 
this relation to express $V_{\it ext}$ in terms of the density when
solving for the spin current $\vec{w}$. We get 
\be 
 n\vec{w} = -\frac{\tau T}{m}\left( \vec{\nabla} M 
  - \frac{M}{n}\vec{\nabla n} \right) \, ,
\ee
in agreement with equ.~(\ref{Fick_2}). 

 In the opposite limit, that of infinite collision time, we expect 
the spin hydrodynamic equations to agree with solutions of the ballistic
Boltzmann equation. In a trap these solutions correspond to simple
spin-sloshing modes. Consider 
\bea 
f_{p\sigma} (x,t) &=& n_0(x_\perp,p_\perp) \exp\left( - \frac{m\omega_z^2}{2T} 
   \left[ z-\bar\sigma z_0\cos(\omega t)\right]^2 \right) \nonumber \\
 && \hspace{1.2cm}\times
\exp\left( - \frac{1}{2mT} 
   \left[ p_z-\bar\sigma p_0\sin(\omega t)\right]^2 \right)
\eea
with $\bar\sigma=\pm$ for $\sigma=\uparrow\downarrow$ and 
\be 
n_0(x_\perp,p_\perp)=\exp\left( - \frac{m\omega_\perp^2 x_\perp^2}{2T} 
-\frac{p_\perp^2}{2mT}   \right)\, . 
\ee
This distribution solves the ballistic Boltzmann equation in a trap
if $\omega=\omega_z$ and $p_0=z_0m\omega_z$. We can compute the spin
densities
\be 
\label{n_slosh}
 n_\sigma = n_0 \exp\left( - \frac{m\omega_z^2}{2T} 
   \left[ z-\sigma z_0\cos(\omega t)\right]^2 \right)
\ee
and the spin velocity $\vec{u}_\sigma= \pm \vec{w}$ with $w_{z} = p_0/m = 
\omega_zz_0$. The spin stresses are given by 
\be 
\label{Pi_slosh}
 \Pi^{ij}_\sigma = mn_\sigma w^iw^j + n_\sigma T\delta^{ij}\, . 
\ee 
It is now straightforward to check that equ.~(\ref{n_slosh}-\ref{Pi_slosh})
satisfies the spin continuity equations (\ref{shydro_0}) and the spin
Euler equation
\be 
\label{shydro_e}
\partial_0 \left( mn_\sigma u^i_\sigma \right)
  + \nabla_j \Pi_\sigma^{ij} =- mn_\sigma F^i \, .
\ee
It is then reasonable to assume that spin hydrodynamics can describe
the transition between diffusion and spin oscillations in a trap.

\section{Simulating spin hydrodynamics}
\label{sec_sim}

 We have implemented spin hydrodynamics in close analogy with our 
implementation of viscous fluid dynamics \cite{Schafer:2010dv}
and anisotropic fluid dynamics \cite{Bluhm:2015raa} for cold atomic
Fermi gases. The numerical code is based on the PPM (piecewise parabolic 
method, Lagrangian remap) method of Colella and Woodward \cite{Colella:1984},
as implemented in the VH1 code developed by Blondin and Lufkin
\cite{Blondin:1993}. We solve the conservation laws using Lagrangian
coordinates. The momentum equations can be written as 
\be 
 D_\sigma u^i_\sigma = -\frac{1}{\rho_\sigma}\, \nabla^i P_\sigma 
  \mp \frac{\rho_g}{\rho_\sigma\tau} \, w^i\, , 
\ee
where $D_\sigma = \partial_0 + \vec{u}_\sigma \cdot\vec\nabla$ is the 
comoving derivative, $\rho_\sigma=mn_\sigma$ is the mass density, and 
$P_\sigma=n_\sigma T$ is the partial pressure of the spin state $\sigma$. 
After a Lagrangian time step the hydrodynamic quantities are remapped 
onto an Eulerian grid. The spin current $\vec\jmath_M = M\vec{u}+n\vec{w}$ 
can be compared to the expectation from Fick's law,  $\vec\jmath_M=
M\vec{u}-D\vec\nabla M$, where $D=\tau T/m$. 

 We consider diffusion in an axially symmetric trapping potential 
$V(x)=\frac{1}{2}m\omega_i^2x_i^2$ with $\omega_x=\omega_y=\omega_\perp$ 
and $\omega_z=\lambda\omega_\perp$. We introduce dimensionless variables 
for distance, time and velocity based on the following system of 
units~\cite{Schafer:2010dv}
\be
\label{x_0_def}
 x_0 = (3N\lambda)^{1/6}\left(\frac{2}{3m\omega_\perp}\right)^{1/2} \, , 
 \hspace{0.5cm}
 t_0 = \omega_\perp^{-1} \, ,
 \hspace{0.5cm}
 u_0 = x_0\omega_\perp \, ,
\ee
where $N=N_\uparrow+N_\downarrow$ is the total number of particles. The unit 
of density is $n_0=x_0^{-3}$, and the unit of temperature is $T_0 = m
\omega_\perp^2x_0^2$. Finally, the unit of the diffusion constant in 
\be 
 D_0 = \omega_\perp x_0^2\, . 
\ee
We will use an overbar to denote dimensionless quantities, for 
example $\bar{x}=x/x_0$, $\bar{T}=T/T_0$, and $\bar{D}=D/D_0$. 

In the high temperature limit the initial density is a Gaussian. The
density is 
\be
n(x) = n(0)\exp\left(-\frac{E_F}{E_0}
  \left[ \bar{x}^2+\bar{y}^2+\lambda^2\bar{z}^2 \right] \right) \, ,
\ee
where $\bar{x}=x/x_0$ is the dimensionless position, $E_F=(3N\lambda)^{1/3}
\omega_\perp$ is the Fermi energy in the trap, and $E_0$ is the total 
energy per particle of the trapped gas. For an ideal gas $E_0=3NT$, and
the dimensionless temperature is $\bar{T}=\frac{1}{2}(E_0/E_F)$. The 
central density is given by 
\be 
\label{n_0}
 n(0) = n_0 \frac{N\lambda}{\pi^{3/2}}\left(\frac{E_F}{E_0}\right)^{3/2} \, .
\ee
It is convenient to normalize the central density to one \footnote{
In this work we will focus on the high temperature limit, so that the 
equilibrium density is a Gaussian. We note, however, that the choice
of units $\bar{n}=n/n_{\it id}(0)$, where $n_{\it id}(0)$ is the central
density of the ideal gas at the same temperature is convenient also
for a general equation of state.}, so that $\bar{n}=n/n(0)$ and 
$\bar{M}=M/n(0)$. 

 A simple parameterization of the diffusion constant can be given 
in terms of a density independent part, reflecting the low temperature
(quantum) behavior, and a part that scales inversely with density, 
corresponding to the high temperature (kinetic) limit. We write
\be 
\label{D_par}
 D = \frac{\beta}{m} + \frac{\beta_T}{m} \frac{(mT)^{3/2}}{n}\, , 
\ee
where $\beta$ and $\beta_T$ are constants. The kinetic theory result
given in equ.~(\ref{D_CE}) corresponds to $\beta_T=3/(16\sqrt{\pi})$.
In dimensionless units this formula becomes
\be 
\label{D_dim}
 \bar{D} = \bar\beta + \bar\beta_T \frac{\bar{T}^{3/2}}{\bar{n}}\, 
\ee
where $\bar{D}=D/D_0$ and 
\be 
\label{beta_dim}
 \bar\beta = \frac{3}{2}\frac{\beta}{(3\lambda N)^{1/3}}\, ,\hspace{0.5cm}
 \bar\beta_T = \frac{4\pi^{3/2}}{3}\frac{\beta_T}{(3\lambda N)^{1/3}}
  \left(\frac{E_0}{E_F}\right)^{3/2}\, . 
\ee
Using these parameters we can provide some simple estimates for the 
time scales involved in simulations of diffusion in a trapped atomic gas. 
We saw that empirically the spin decay rate scales as $\Gamma=\omega_z^2
/(\gamma E_F)\cdot (T/T_F)^{1/2}$, see the discussion preceding 
equ.~(\ref{G_sd}). The experiment of Sommer et al.~gives $\gamma\simeq
0.16$. Based on the units described above the dimensionless decay time
is  
\be 
\label{diff_scale}
 \bar\Gamma^{-1} = 2.87\gamma 
     \frac{(\lambda N)^{1/3}}{\lambda}
     \left( \frac{E_F}{E_0} \right)^{2}\, . 
\ee
where $\bar\Gamma=\Gamma/\omega_\perp$. Sommer et al.~do not provide the 
precise values of $\lambda$ and $N$ in their experiment, but typical values 
used in the viscosity measurements reported in \cite{Kinast:2005,Cao:2010wa} 
are $N=2\cdot 10^5$ and $\lambda=0.045$. These parameters lead to long decay
times $\bar\Gamma^{-1} \simeq 212 (E_F/E_0)^{2}$.

 This estimate should be compared to the typical time step in a spin
hydrodynamic simulation. In ordinary fluid dynamics the time step 
is controlled by the speed of sound and the resolution, $\Delta t 
=C\Delta x/c_s$, where the Courant number $C$ is typically chosen
to be 1/2. Using dimensionless units and the speed of sound of an 
ideal gas we find
\be 
\label{cs_scale}
 \Delta\bar{t} = C\sqrt{\frac{6}{5}} \left(\frac{E_F}{E_0}\right)^{1/2}
 \Delta\bar{x}\, . 
\ee
The units are chosen such that the cloud size is of order 1. Then $\Delta
\bar{t}\lsim \Delta x \lsim 0.1$ is a typical time step for the hydrodynamic
evolution. In spin fluid dynamics we also have to ensure that the time step 
is small compared to the relaxation time. The dimensionless relaxation time is 
\be 
\label{relax_scale}
\bar\tau = \frac{\bar\beta}{\bar{T}} 
   + \frac{\bar\beta_T \bar{T}^{1/2}}{\bar{n}}\, . 
\ee
Using the estimate $\beta_T=3/(16\sqrt{\pi})$ together with 
equ.~(\ref{beta_dim}), as well as the values of $N$ and $\lambda$ 
given above, we get $\bar\tau(0) = 0.02 (E_0/E_F)^{2}$. This suggests that 
for small $\lambda$ and typical values of $E_0/E_F$ there is a significant
disparity of scales between the diffusive scale equ.~(\ref{diff_scale}) 
and the relaxation scale equ.~(\ref{relax_scale}). As a result, in the 
limit that the cloud is very deformed ($\lambda\to 0$) and the diffusion 
constant is very small ($\bar\beta\to 0$), spin hydrodynamics is 
potentially an inefficient method for simulating the diffusion equation. 
This is not necessarily a problem. First, if the diffusion constant is 
small diffusive behavior sets in quickly and the decay constant can 
be accurately determined even if the simulation time is less that 
$\Gamma^{-1}$. Second, a similar disparity of scales appears in the 
anisotropic hydrodynamics method as the shear viscosity becomes small. 
Anisotropic hydrodynamics is indeed an inefficient method for solving 
the Euler equation, but a powerful tool to extract the shear viscosity 
for realistic geometries \cite{Bluhm:2015bzi}.

\section{Numerical results: Box}
\label{sec_num}

\begin{figure}[t]\bc
\includegraphics[width=7.75cm]{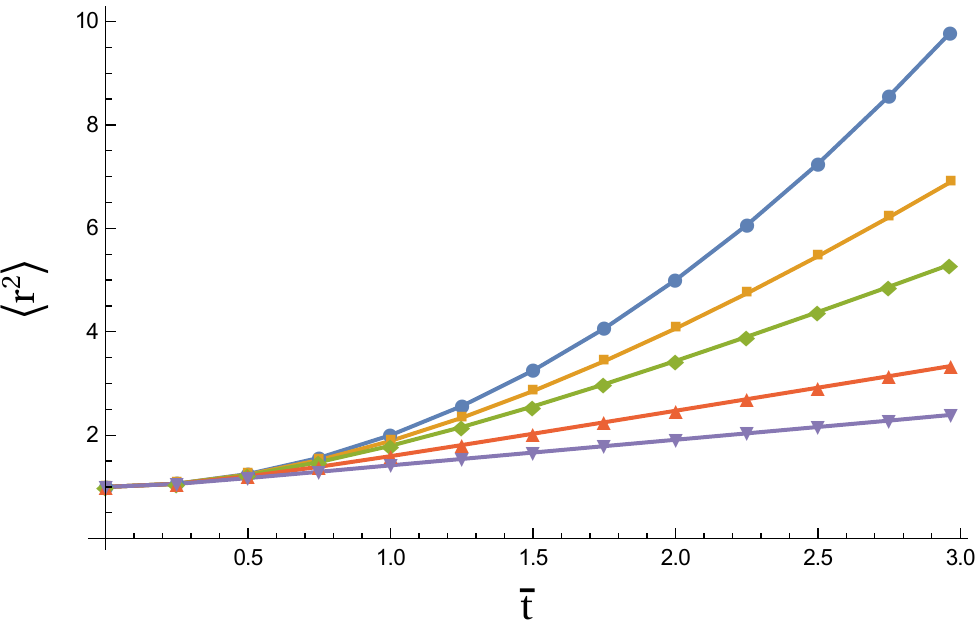}
\vspace*{0.2cm}
\includegraphics[width=7.75cm]{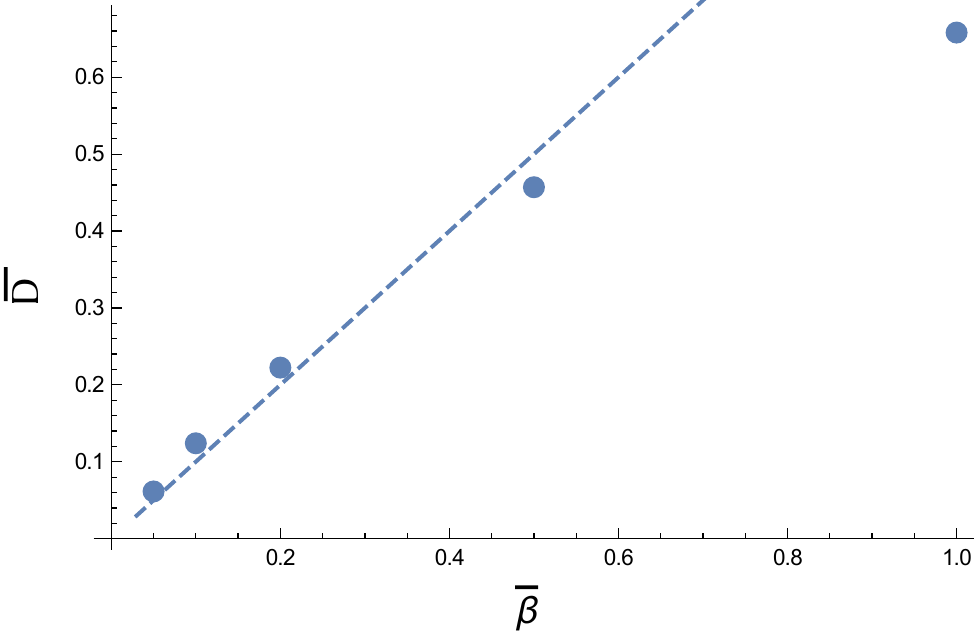}\ec
\caption{\label{fig_dif}
The left panel shows the mean square size $\langle r^2\rangle$ of the 
magnetization $M=n_\uparrow-n_\downarrow$ as a function of time for the 
evolution of a Gaussian initial state. The different curves correspond 
to different values of the diffusion parameter, from top to bottom 
$\bar\beta=(1000,1,0.5,0.2,0.1)$. We observe the transition from 
free expansion, $\langle r^2\rangle \sim \bar{t}^2$, to diffusion,
$\langle r^2\rangle \sim \bar{t}$. The right panel shows the diffusion
constant extracted from the growth of $\langle r^2\rangle$. The dashed curve
shows the theoretical expectation in the small $\bar\beta$ limit.}   
\end{figure}

\begin{figure}[t]\bc
\includegraphics[width=7.75cm]{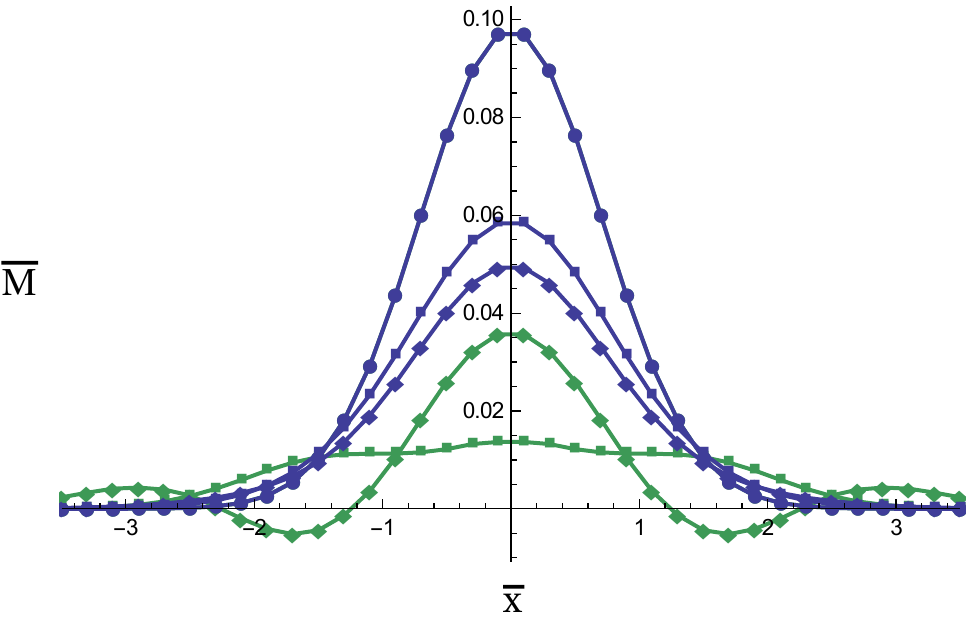}
\vspace*{0.2cm}
\includegraphics[width=7.75cm]{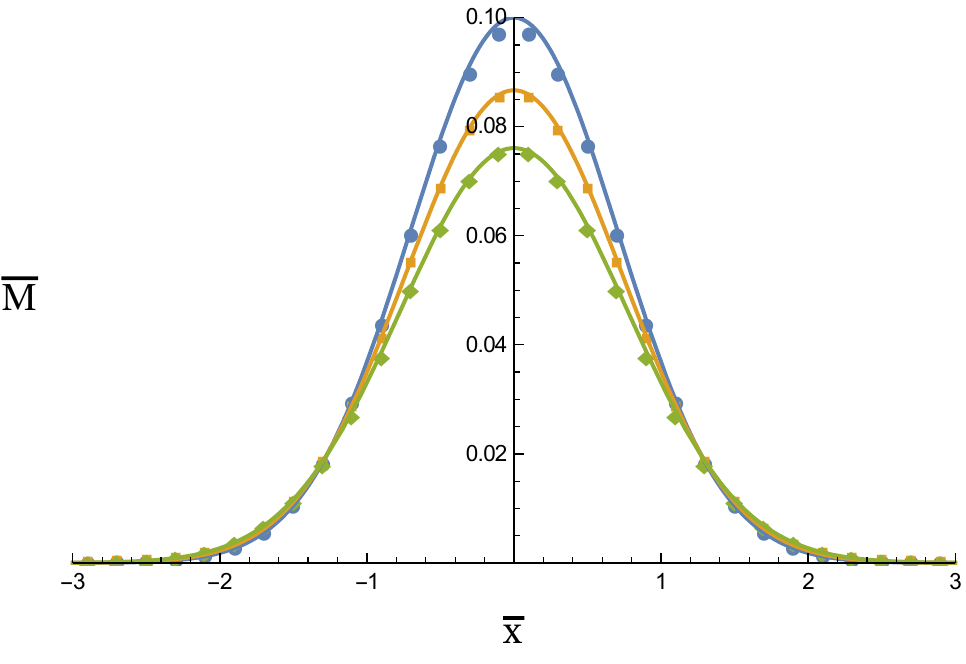}
\ec
\caption{\label{fig_m}
The left panel shows the time evolution of the dimensionless magnetization 
$\bar{M}(\bar{x},\bar{t})$ for two different values of $\bar\beta=1000$ 
(green diamonds) and $\bar\beta=0.1$ (blue circles). The curves at $\bar{t}
=0$ (top) are identical, and only the $\bar\beta=0.1$ graph is visible. The 
time step between successive curves is $\Delta\bar{t}=1.25$. The right panel 
shows the time evolution of $M(\bar{x},\bar{t})$ for a small value of 
$\bar\beta=0.05$. The dots show the result of spin hydrodynamics at different 
time steps separated by $\Delta\bar{t}=0.5$ (time increasing from top to 
bottom), and the lines are the expectations from the diffusion equation 
(\ref{d_gauss}).  }   
\end{figure}

 In order to test spin hydrodynamics we have solved the equations
of motion in a three-dimensional box. The simulation is carried
out on a three dimensional cartesian grid with $50^3$ points and 
a grid spacing $\Delta\bar{x}=0.2$. We consider a constant background
density $\bar{n}_\uparrow=\bar{n}_\downarrow=1/2$ with a Gaussian 
perturbation $\delta\bar{n}_{\uparrow\downarrow}=\pm 0.05\exp(-\bar{x}_i^2)$.
The left panel in Fig.~\ref{fig_m} shows the evolution of the mean
square magnetization radius
\be
 \langle r^2\rangle = \frac{1}{M_{\it tot}}
\int d^3\bar{x} \, \bar{x}^{2}_i M(\bar{x},\bar{t})
\ee 
as a function of time. Here, $M_{\it tot}$ is the integrated magnetization. The 
plot shows the result for a range of values of $\bar\beta$, corresponding to 
a range of relaxation times. We note that in a box, in which the background 
density is constant, there is no difference between the scaling with 
$\bar\beta$ and $\bar\beta_T$. In the limit of large $\bar\beta$ the squared 
radius grows quadratically with time, corresponding to a constant spin 
velocity $\vec{w}$ and ballistic expansion. For small values of $\bar\beta$ 
the squared radius grows linear with time, as expected from the solution of 
the diffusion equation. The diffusion equation predicts
\be 
\label{d_gauss}
 M(\bar{x},\bar{t}) = \frac{M_0}{(1+4\bar{D}\bar{t})^{3/2}}\, 
 \exp\left(-\frac{\bar{x}^2}{1+4\bar{D}\bar{t}}\right)\, .
\ee
In the right panel of Fig.~\ref{fig_dif} we show the diffusion constant
extracted from the slope of $\langle r^2\rangle$ together with the 
theoretical expectation $\bar{D}=\bar\beta$. The agreement for small
$\bar\beta$ is quite good. In this regime there is a systematic shift 
between $\bar\beta$ and the extracted value of $D$, which indicates
some amount of numerical diffusion.

  In Fig.~\ref{fig_m} we show the evolution of the magnetization in more 
detail. The left panel of Fig.~\ref{fig_m} demonstrates that for large 
$\bar\beta$ (large relaxation time) the evolution is not diffusive. There 
is a magnetization front which propagates at approximately constant
speed. For small $\bar\beta$ (small relaxation time), on the other hand, 
the evolution is consistent with diffusion. This is seen more clearly in 
the right panel of Fig.~\ref{fig_m}, in which we compare the time and 
spatial dependence of the magnetization in spin hydrodynamics with the 
prediction from the diffusion law in equ.~(\ref{d_gauss}).

\begin{figure}[t]\bc
\includegraphics[width=9.5cm]{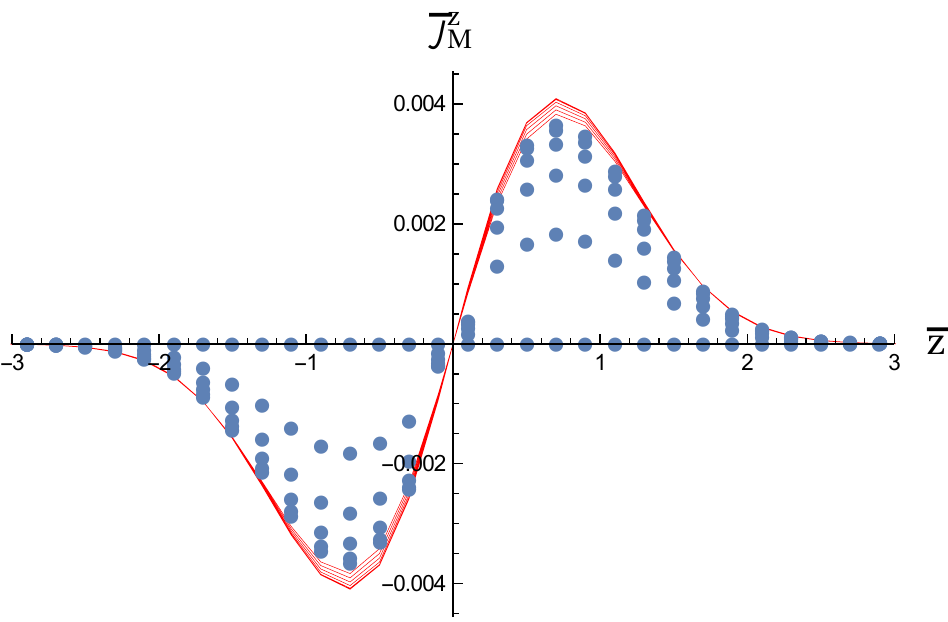}
\ec
\caption{\label{fig_cur}
Spin current $\vec\jmath_M=n\vec{w}+M\vec{u}$ in spin hydrodynamics (dots)
compared to the expectation from Fick's law, $\vec\jmath_M=-D\vec\nabla M$ 
(lines). We show the $z$-component of the dimensionless current as a function 
of $\bar{z}$ (with $\bar{x}=\bar{y}=0$) for $\bar\beta=0.05$ and several 
values of $\bar{t}=(0,0.05,0.10,0.15,0.20)$. Note that the prediction from 
Fick's law starts maximal and then decays (very slowly, on the time scale 
shown in this figure), whereas the current in spin hydrodynamics starts at 
zero and the approaches Fick's law. }   
\end{figure}

 In Fig.~\ref{fig_cur} we compare the spin current $\jmath_M$ in 
spin hydrodynamics with the expectation from Fick's law, $\vec\jmath_M=-
D\vec\nabla M$. Note that in the present case there is no convective 
contribution $M\vec{u}$. Fick's law predicts that the spin current
turns on instantaneously, and then decays slowly as the cloud 
expands. Spin hydrodynamics, on the other hand, predicts that the
spin current vanishes at $\bar{t}=0$ and then approaches Fick's
law on a time scale set by the relaxation time. At late time the 
spin hydrodynamics current tracks Fick's law. 

\section{Numerical results: Trapped gas}
\label{sec_num_trap}

\begin{figure}[t]\bc
\includegraphics[width=9.5cm]{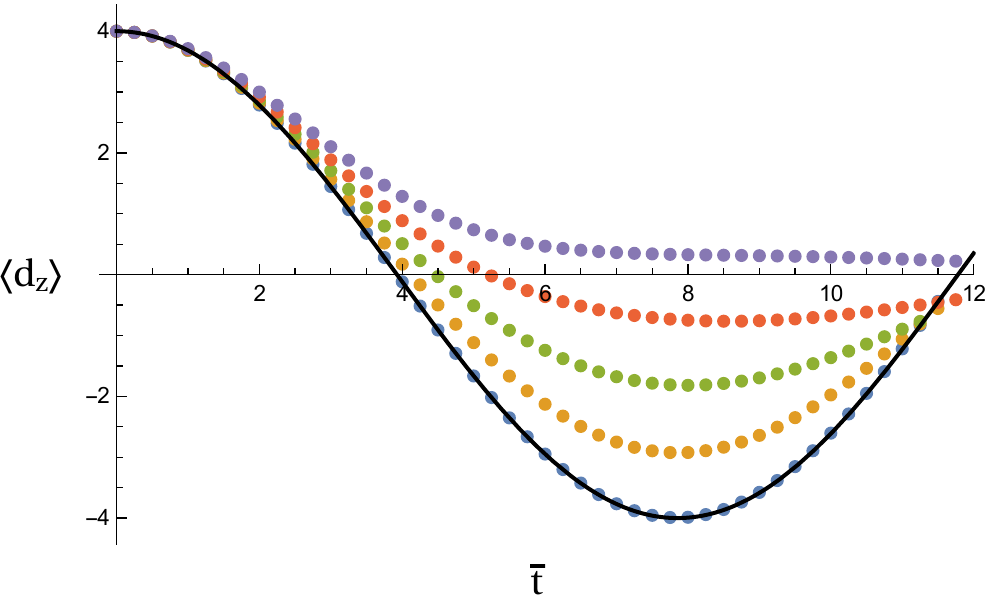}
\ec
\caption{\label{fig_dip}
Evolution of the spin dipole moment in a trapped gas a function of 
time. The initial condition is given by two shifted Gaussians, see
equ.~(\ref{init_gauss}). The solid line shows an undamped spin oscillation 
with frequency $\bar\omega=0.4$. The points show the results of a 
spin hydrodynamics simulation with $\bar\beta=(1000,5,2,1,0.5)$,
going from oscillatory to overdamped behavior. }   
\end{figure}

 In this section we will consider a harmonically trapped gas. 
We assume axial symmetry, and the simulations are carried out 
in cylindrical coordinates on a grid with dimensions $50^2$ and
grid spacing $\Delta\bar{z}=0.2$ and $\Delta\bar\rho=0.2$. The
main observable is the spin dipole moment 
\be 
 d_z = \frac{2}{N_{\it tot}}\int d^3\bar{x} \, 
       \bar{z}\, M(\bar{x},\bar{t})\, ,
\ee
which is the same quantity that was studied in the experimental work 
of Sommer et al.~\cite{Sommer:2011}. We first consider a density 
independent relaxation time, governed by the parameter $\bar\beta$.
The initial spin density is given by two shifted Gaussians
\be 
\label{init_gauss}
 \bar{n}_\sigma = \frac{1}{2}\exp\left(-\frac{E_F}{E_0}
  \left[\lambda^2(\bar{z}\pm \bar{z}_0)^2+\bar\rho^2\right]
   \right).
\ee
We use $E_0/E_F=1$, $\lambda=0.4$ and $\bar{z}_0=2$. For $\bar\beta\to
\infty$ we expect the system to show undamped spin oscillations with 
frequency $\bar\omega=\lambda$, as described in Sect.~\ref{sec_lim}.
This can be seen in Fig.~\ref{fig_dip}. For finite but large $\bar
\beta$ the gas exhibits damped oscillations, and for small $\beta$
the motion is overdamped. 

\begin{figure}[t]\bc
\includegraphics[width=7.75cm]{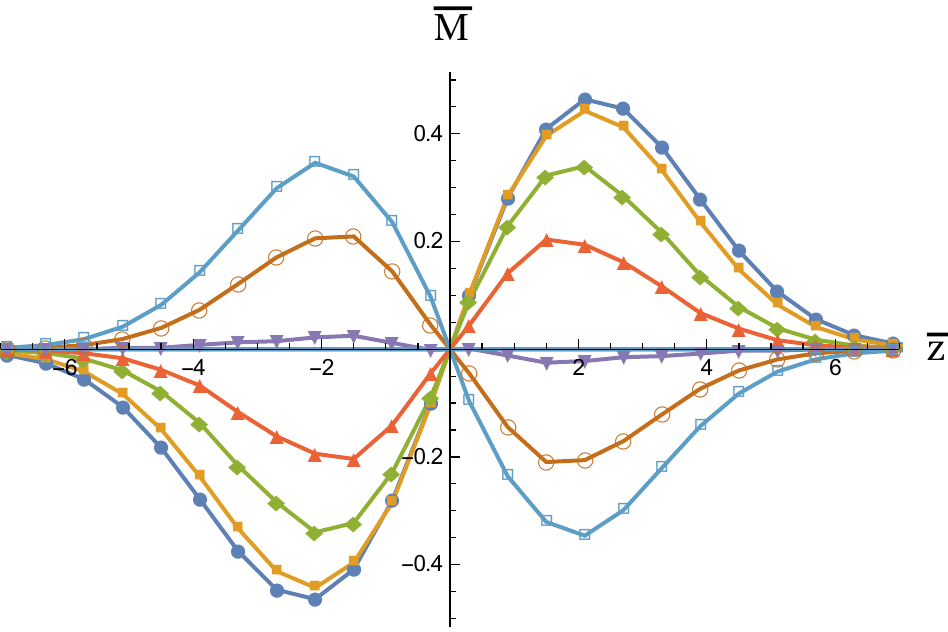}
\vspace*{0.35cm}
\includegraphics[width=7.75cm]{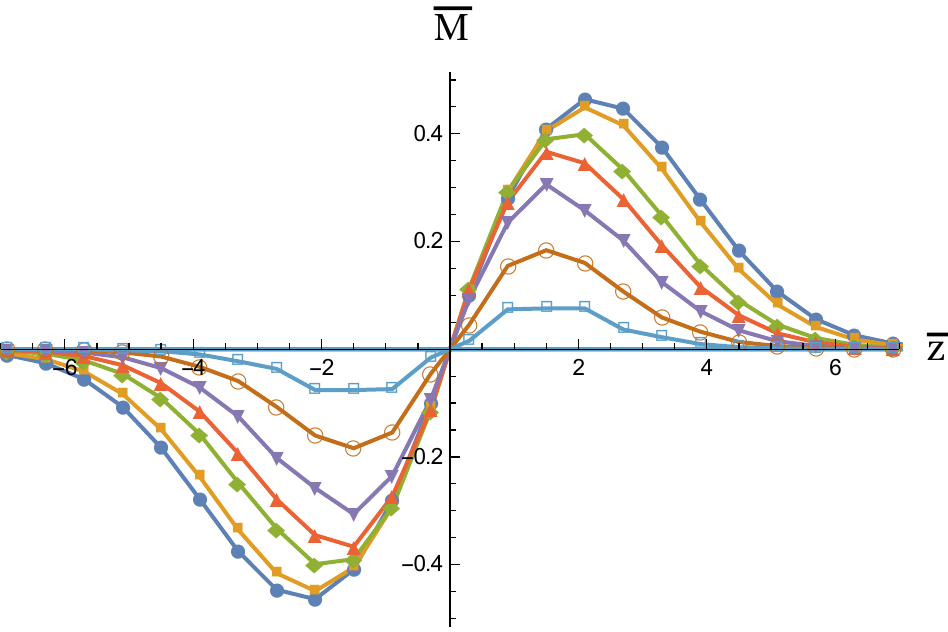}
\ec
\caption{\label{fig_dip_M}
Magnetization as a function of position for a trapped Fermi gas. The 
left panel shows the magnetization for different times in the ballistic
(spin oscillation) limit $\bar\beta=1000$. The curves are separated by 
$\Delta\bar{t}=1.25$, starting with $\bar{t}=0$ (blue circles). The right 
panel shows the magnetization at different times for $\bar\beta=1$, closer 
to the diffusive limit.  }   
\end{figure}

 More details are shown in Fig.~\ref{fig_dip_M}. The left and right 
panels shows the evolution of the magnetization for $\bar\beta=1000$
and $\bar\beta=1$, respectively. We observe that for $\bar\beta=1000$
the magnetization oscillates, and for $\bar\beta=1$ it is strictly
decaying.  The decay is not precisely exponential, because the decay
of the magnetization is superimposed on an undamped quadrupole 
oscillation of the total density. Physically, this mode is damped
by shear viscosity, but we have not included viscosity in our study. 
Another possibility is to consider initial conditions that correspond
to the late time dynamics of the trapped gas, and for which the total
density is equilibrated. We choose
\be 
\label{init_dip}
 \bar{n}_\sigma = \frac{1}{2}\left( 
  1\pm A\frac{\bar{z}}{1+\lambda^2\bar{z}^2+\bar\rho^2}\right) 
\exp\left(-\frac{E_F}{E_0}
  \left[\lambda^2\bar{z}^2+\bar\rho^2\right]
   \right)\, , 
\ee
which is motivated by the variational results derived in Sect.~\ref{sec_sol}.

\begin{figure}[t]\bc
\includegraphics[width=7.75cm]{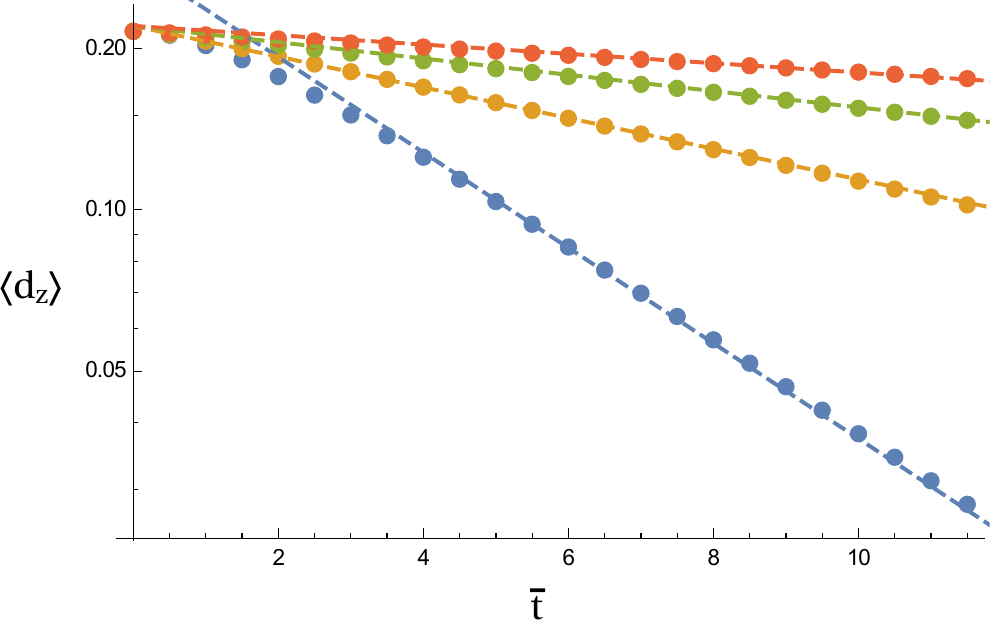}
\includegraphics[width=7.75cm]{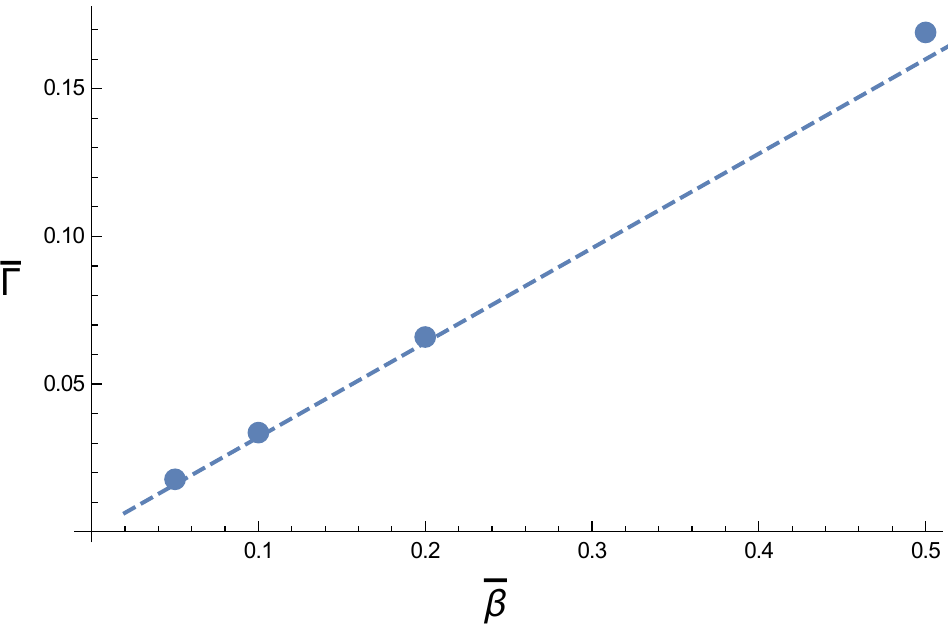}
\ec
\caption{\label{fig_dip_i}
The left panel shows the time evolution of the spin dipole moment in a 
trapped gas with a density independent diffusion constant. The initial 
condition is given by equ.~(\ref{init_dip}). The points show the results 
of a spin hydrodynamics simulation with $\bar\beta=(0.5,0.2,0.1,0.05)$, 
and the dashed lines are exponential fits. The right panel shows the 
extracted spin decay constant $\bar\Gamma$ as a function of $\bar\beta$. 
The dashed line corresponds to $\Gamma_{\it red}=2$ in equ.~(\ref{beta_G}). }   
\end{figure}

 The evolution of the spin dipole moment is shown in Fig.~\ref{fig_dip_i}.
The left panel demonstrates that the decay of the dipole moment is indeed 
exponential. The right panel shows the dependence of the decay constant
on $\bar\beta$. For small $\bar\beta$ we observe a linear relationship. 
This behavior can be compared with the solution of the diffusion 
equation obtained in Sect.~\ref{sec_sol}. We obtained $\Gamma=\frac{D_0}
{l_z^2}\Gamma_{\it red}$ with $\Gamma_{\it red}=2$. In dimensionless units
this can be written as
\be 
\label{beta_G}
 \bar\Gamma = \frac{1}{2\bar{T}}\,\bar\beta\lambda^2 \, \Gamma_{\it red}\, . 
\ee 
This relation is shown as the dashed line in the right panel of 
Fig.~\ref{fig_dip_i}. We observe that $\Gamma_{\it red}=2$ indeed provides
a very good description of the data for $\bar\beta\lsim 0.5$. We 
conclude that spin hydrodynamics indeed converges to the expected 
solution of the diffusion equation in a trapped geometry. 

 We are now in a position to study the problem that motivated this
study. Consider a diffusion constant which is inversely proportional
to density, governed by the parameter $\bar\beta_T$ in 
equ.~(\ref{D_par},\ref{D_dim}). We study the evolution in a deformed
trap, beginning from the initial condition given in equ.~(\ref{init_dip}).
As explained in Sect.~\ref{sec_sol} the diffusion equation predicts 
that for fixed diffusion constant $D_0$ at the trap center the decay
of the spin polarization is much faster. This effect is caused by 
a large spin current in the dilute regime. In spin hydrodynamics, on
the other hand, the relaxation time in the dilute regime is large,
and we do not expect a large spin current to develop. 

\begin{figure}[t]\bc
\includegraphics[width=7.75cm]{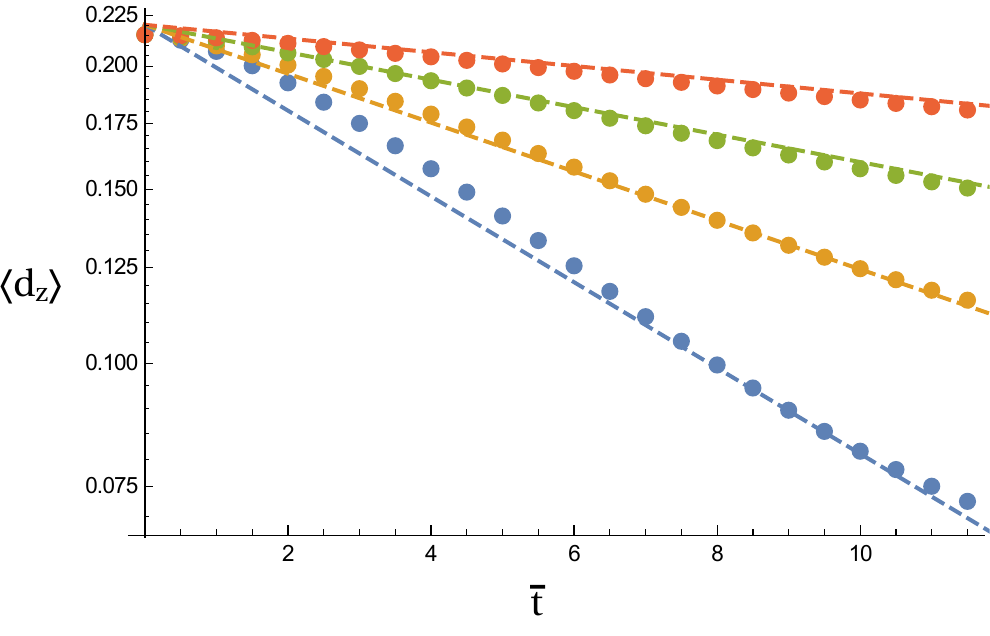}
\vspace*{0.2cm}
\includegraphics[width=7.75cm]{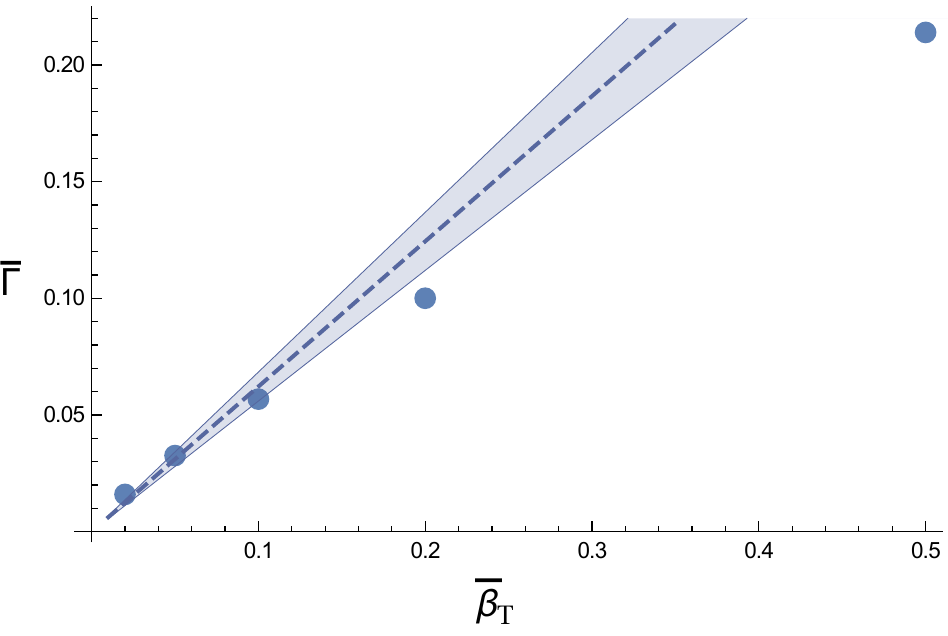}
\ec
\caption{\label{fig_dip_bt}
The left panel shows the time evolution of the spin dipole moment in a 
trapped gas with $D\sim 1/n$. The initial condition is given by 
equ.~(\ref{init_dip}). The points show the results of a spin hydrodynamics 
simulation with $\bar\beta_T=(0.2,0.1,0.05,0.02)$, and the dashed lines are 
exponential fits. The right panel shows the extracted spin decay constant 
$\bar\Gamma$ as a function of $\bar\beta$. The dashed line corresponds to 
$\Gamma_{\it red}=11$ in equ.~(\ref{beta_T_G}). The band shows a $\pm 10\%$
uncertainty in $\Gamma_{\it red}$. }   
\end{figure}

 The time evolution of the spin dipole moment for different values of
$\bar\beta_T$ is shown in the left panel of Fig.~\ref{fig_dip_bt}. 
We observe that for $\beta_T\lsim 0.2$ the decay of the spin polarization 
is exponential. The extracted spin decay constant is shown in the right
panel of Fig.~\ref{fig_dip_bt}. As before, we can compare the result
to solutions of the diffusion equation. In dimensionless units we
get 
\be 
\label{beta_T_G}
 \bar\Gamma = \frac{1}{2\bar{T}}\,\bar\beta\lambda^2\bar{T}^{3/2} 
     \, \Gamma_{\it red}\, . 
\ee 
We found that the diffusion equation predicts $\Gamma_{\it red}(0.4)
=22.9$, whereas the experiment of Sommer et al.~\cite{Sommer:2011}
indicates that $\Gamma_{\it red}=11.3$. Note that this result assumes
the validity of kinetic theory, in particular the relation $D(0)=0.106 
(mT)^{3/2}/(mn(0))$, see equ.~(\ref{D_CE}). In spin hydrodynamics we can 
extract $\Gamma_{\it red}$ from the slope of the $\bar\beta_T-\bar\Gamma$ 
relation. The dashed line in the right panel of Fig.~\ref{fig_dip_bt} 
corresponds to $\Gamma_{\it red}=11$, and the error band indicates that 
the uncertainty in this analysis is about 10\%. We can therefore 
deduce that  
\be
 D(0) = (0.1\pm 0.01)\times\frac{(mT)^{3/2}}{mn(0)}\, . 
\ee
As a consistency check we have studied the dependence on the trap 
deformation $\lambda$. We have repeated the analysis shown in 
Fig.~\ref{fig_dip_bt} for a smaller value $\lambda=0.25$. We 
find smaller decay constants $\bar\Gamma$, and a slightly delayed
onset of the linear behavior in the $\bar\Gamma-\bar\beta_T$ plot, 
but the reduced decay constant $\Gamma_{\it red}=11\pm 1$ is 
unchanged. This is consistent with the experimental finding that 
the reduced decay constant does not depend on the trap deformation. 

We note that the linear scaling with $\bar\beta_T$ implies that 
the damping constant is proportional to $\bar{T}^{3/2}E_0^{3/2}\sim T^3$. 
The first factor arises from the temperature dependence of the 
diffusion constant, and the second factor is due to the relation 
$T_F(0)\sim T^{-1}$ at fixed $N$ and $\omega_\perp,\omega_z$. The overall
scaling of the damping constant contains an extra factor $l_z^{-2}
\sim T^{-1}$, so that $\Gamma\sim T^2$. This is indeed the behavior
observed in \cite{Sommer:2011}.
 
\section{Conclusions and outlook}
\label{sec_out}

 In this work we have derived the equations of spin hydrodynamics
from an underlying kinetic theory. Spin hydrodynamics reduces to the 
diffusion equation in the dense limit, and to ballistic motion in
the dilute limit. We have validated a numerical implementation of 
spin hydrodynamics using a number of test cases. The diffusive limit
was studied using the expansion of a Gaussian magnetization in a gas 
at constant density, and by following the decay of the spin dipole 
mode in a harmonic trap with density independent diffusion constant. 
The ballistic limit was studied using the spin slosh mode in a 
harmonic trap. 

 We applied spin hydrodynamics to the decay of the spin dipole 
mode in a dilute Fermi gas at unitarity. In the high temperature
limit kinetic theory predicts that $D\sim T^{3/2}/n$. We verified 
that the experiment of Sommer at al.~\cite{Sommer:2011} is consistent 
with this prediction, and that the coefficient of proportionality 
agrees with kinetic theory. This conclusion was previously reached 
in the beautiful work of Bruun and Pethick \cite{Bruun:2011b}, but 
these authors were forced to introduce an unknown parameter, the 
radial cutoff in the diffusion equation. Our method has no free 
parameters other than the diffusion constant. Sommer et al.~concluded
that agreement with kinetic theory can be achieved if the diffusion
constant is corrected for the finite size of the trap. 

\begin{figure}[t]\bc
\includegraphics[width=7.75cm]{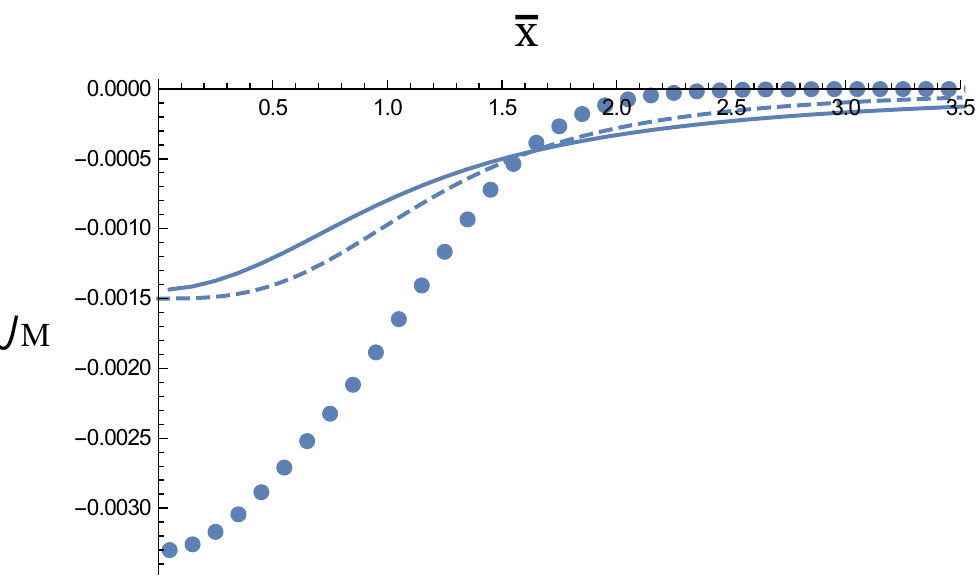}
\vspace*{0.2cm}
\includegraphics[width=7.75cm]{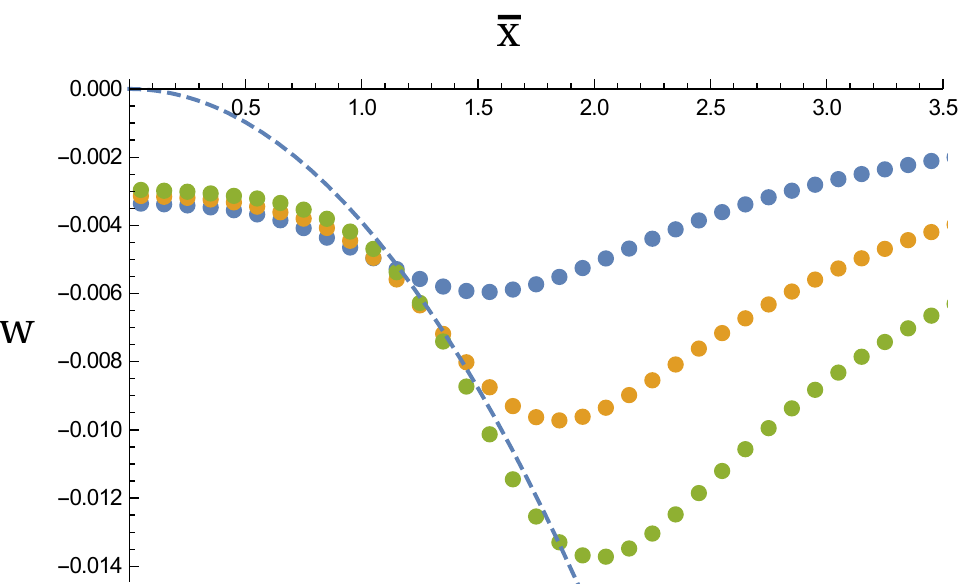}
\ec
\caption{\label{fig_cur_t}
Longitudinal spin current $\jmath_M$ (left panel) and spin velocity $w$
(right panel) in the transverse plane. We show the $z$-component of the 
current and the velocity at $\bar{z}=0$ as a function of the transverse 
position $\bar{x}$ for $\bar\beta_T=0.05$. The dots in the left panel show 
the spin current at $\bar{t}=0.25$. The solid line is the expectation from 
Fick's law, and the dashed line is the variational estimate of the current 
profile obtained in Sect.~\ref{sec_sol} (scaled to fit Fick's law). The
right panel shows the spin current at different times $\bar{t}=0.25,0.50,
0.75$ (top to bottom). The dashed line is the variational estimate of 
the drift velocity from \cite{Sommer:2011}, scaled to fit the data. }   
\end{figure}

 A more detailed comparison to earlier work is shown in 
Fig.~\ref{fig_cur_t}. The figure displays the profile of the spin current 
$\jmath_M$ and the spin velocity $w$ in the transverse plane. We consider
a diffusion constant of the form $D\sim T^{3/2}/n$, and we choose $\bar
\beta_T=0.05$. The left panel shows the spin current (dots) compared
to the expectation from Fick's law (solid line) and the variational
estimate discussed in Sect.~\ref{sec_sol}. We observe that the variational
estimate is indeed close to Fick's law, but that the full spin current
is significantly smaller than the variational result for $\bar{x}\gsim 2$. 
This is consistent with the conclusion of Bruun and Pethick that in
order to match experimental data one has to impose a cutoff $r_0\simeq
2.1 l_x$. The right panel shows the spin velocity at different times
$\bar{t}=0.25,0.50,0.75$. For comparison, we show the variational
ansatz for the the drift velocity $w_z\simeq w_z^0 (x/x_0)^2$ proposed
by Sommer et al.~\cite{Sommer:2011}, matched to fit the data. We observe 
that the agreement is very good in the regime $x\gsim l_x$, and that
the data match the variational estimate out to larger distances as
time progresses.  

 Our work can be extended in a number of ways. First, it is important
to further test spin hydrodynamics using detailed comparisons with 
numerical simulations based on the Boltzmann equation in the 
weakly collisional limit. A similar study for anisotropic fluid
dynamics is described in \cite{Bluhm:2015bzi,Pantel:2014jfa}.
Second, we would like to perform precision determinations of 
the spin diffusion constant not only in the high temperature
limit, but also in the vicinity of the critical temperature 
for superfluidity. This will require implementing a more general
functional form of the diffusion constant, and performing detailed
fits of the temperature dependence of the decay rate of the 
spin dipole mode. The ultimate goal of this effort is to provide
determinations of both the shear viscosity and the diffusion constant
in the ``perfect fluid'' regime $a\to\infty$ and $T\sim T_c$, and
to compare the results with expectations from quasi-particle 
theories as well as holographic models 
\cite{Schafer:2009dj,Guo:2010,Adams:2012th}.

 Acknowledgments: This work was supported in parts by the US Department 
of Energy grant DE-FG02-03ER41260. I would like to thank James Joseph
and John Thomas for many useful discussions, and Georg Bruun and Martin 
Zwierlein for comments. I would also like to thank John Blondin for help 
with the VH1 code. This work was completed at the Institute for Nuclear
Theory (INT) in Seattle during the program ``The phases of dense matter''.


\end{document}